# A NOVEL ER MODEL TO RELATIONAL MODEL TRANSFORMATION ALGORITHM FOR SEMANTICALLY CLEAR HIGH QUALITY DATABASE DESIGN

Dhammik Pieris, dhammika.pieris@monash.edu

Monash University

## ABSTRACT

Conceptual modelling using the entity relationship (ER) model has been widely used for database design for a long period of time. However, studies indicate that creating a satisfactory relational model representation from an ER model is uncertain due to the insufficiencies both in the transformation methods used and in the relational model itself. In an effort to solve the issue the original ER notation has been modified, and accordingly, a new transformation algorithm has been developed. This paper presents the proposed transformation algorithm. Using a real world example it shows how the algorithm can be applied in practice. The paper also discusses how to validate the resulted database and reclaim the information that it represents.

## 1. INTRODUCTION

It has been argued that the present "ER to relational transformation algorithm" (Batini, Ceri, & Navathe, 1992; Ramez Elmasri & Navathe, 2007; R Elmasri & Navathe, 2011) is inappropriate for transforming an "ER model"(Chen, 1976) to the "relational model"(Codd, 1970). Our studies revealed that this situation is much more obvious (Pieris & Rajapakse, 2012) when the ER model is presented in its original notation(Chen, 1976). Conversely, the original ER notation has been appreciated for giving a natural view of data that the users want to represent(Cuadra, Martínez, Castro, & Al-Jumaily, 2012). With a view to obtain a successful relational model transformation from an ER model in the original notation, the original ER notation has been modified (Pieris, 2013) and a new transformation algorithm has been designed. This paper presents the new transformation algorithm, an application of it, and a method for validating the resulted database and reclaiming the information represented. A very early version of the algorithm published can be found in (Pieris & Rajapakse, 2012). The current version that this paper presents, is a much more developed one.

The reminder of the paper is as follows. A narrative of a real world scenario and its ER model is presented in section 2. The section 3 provides the background information of the algorithm. The steps of the algorithm and how it can be applied are presented in the next section 4. The section 5 discusses the interpretation of a created database schema (RDS). A conclusion is given in Section 6.

## 2. THE EXAMPLE ER MODEL

Following is a narrative: a description of a company database for creating an ER model. It follows a corresponding ER model given.

> "A company is structured in to departments. Each department has a single name, and a number. Each department is managed by a particular employee. The start date when that employee began managing the department is required to be tracked. A department may be located in several locations. It has a unique number, a unique name, and it also has a field of specialized. Some departments control number of projects. A project has a unique number, a unique name, and a description. An engineer-an employee-is assigned to each project, and it is required recording the date on which an engineer is assigned to a project.
>
> Name, number, address, and the salary are required to be stored for each employee. Dependents of each employee are also recorded for insurance purposes. For each dependent, the name, sex, and relationship to the employee are important to be recorded."

An ER model has been created as given bellow to represent the situation given in the above narrative. It has been drawn following the modified ER notation scheme (Pieris, 2013).

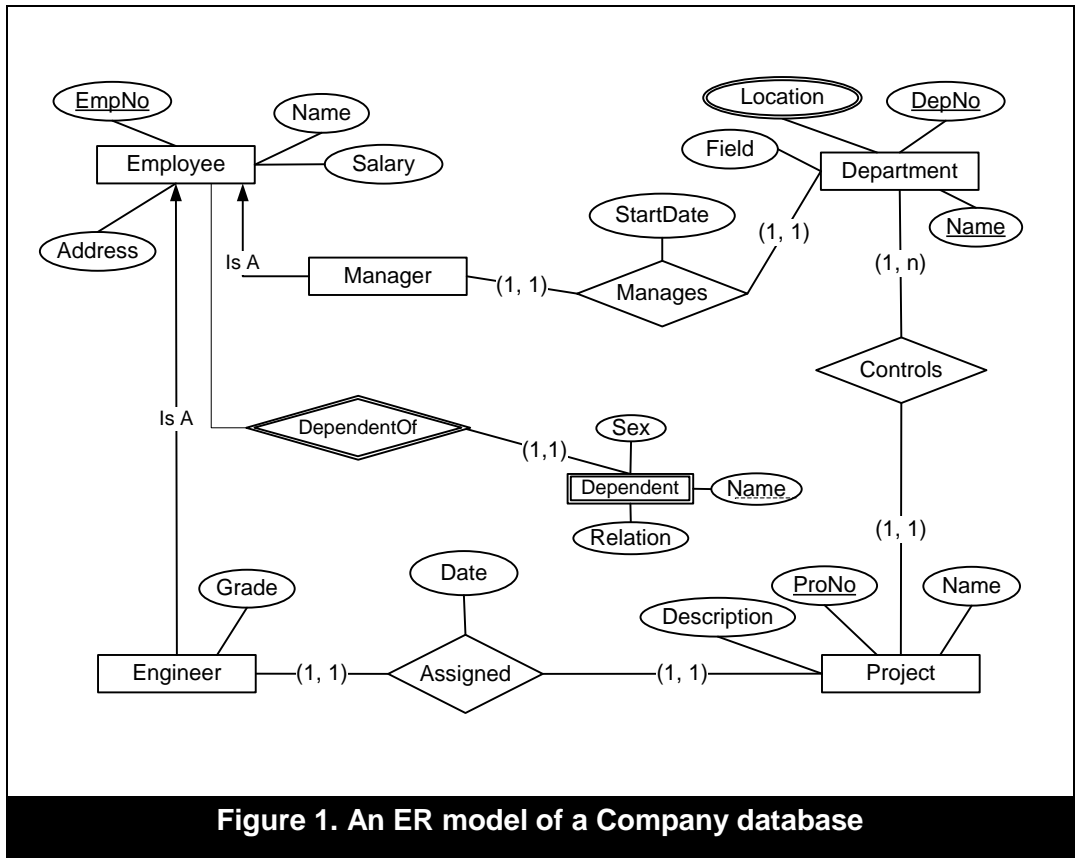

**Figure 1. An ER model of a Company database**

Before transforming any ER model using the new algorithm, it is essential to ensure whether they conform to the new ER model notation scheme(Pieris, 2013). We will set this requirement as a pre requisite for the algorithm. Following description will assist to scrutinize whether the ER model actually conforms to the rules given in the notation.

- A key attribute from each of the regular entity type are named following the rule 2.4.2 in Pieris (2013). For example, "EmpNo" from "Employee", "DepNo" from "Department", and "ProNo" from "Project". Accordingly, the name of each of the key attribute: EmpNo, DepNo, and the ProNo have been made analogy to the name of the respective entity type by making the first three letters of the names of both similar to each other. It has been done by concatenating the first three letters of the name of the respective entity type to the name of the key attribute selected.

- Names of attributes, entity types and relationships are given in singular form (rule 2.5.1) and such names are of capital letter initialized (rule 2.5.1). For example, Location, StartDate, Employee, etc
- Multiple words occurred are concatenated by removing their intermediate spaces (rule 2.5.3), for example, StartDate, DependentOf, etc.
- There are no subtypes that has no attributes and each subtype has at least one intrinsic or a mutual property[1] (rule 2.2.1-I)
- All the names have been freed from including underscores, dashes, hyphens, slashes or any other symbols but alphabetic letters. (rule 2.5.2)

## 3. PREFACE TO THE ALGORITHM

The rules of the algorithm are organized as steps, so that a certain step includes the rules required for transforming a particular part of an given ER model. Accordingly the algorithm contains 14 steps, but only 6 of them are presented in this paper. The selection of the six steps was based on what is minimally required for transforming the ER model given. The rest of the steps will be presented in a paper to be published soon. Steps are named not using numbers but using abbreviations, and each such abbreviation represents the title given to a particular step as given in the Table 1. The last and the 3rd column of the table indicate the title given to the step. The title indicates what is going to be transformed by the respective step. The second column gives a short abbreviation for the title.

---

[1] A mutual property of an entity type means a relationship type that the entity type is participated in

| Table 1: Naming the Steps of the Algorithm | | |
|---|---|---|
| **No** | **Step** | **What to transform** |
| 1 | REG | **Reg**ular entity type |
| 2 | SUB | **Sub**types |
| 3 | GNG | One-to-many (1:**N**) relationship type between two re**g**ular entity type |
| 4 | SOG | **O**ne-to-one (1:1) relationship type between a **s**ubtype and a re**g**ular entity type |
| 5 | MVA | **M**ulti-**v**alued **a**ttribute |
| 6 | WAK | **W**e**ak** entity type |

To transform the ER model, first, apply the Step REG and transform all the available regular entity types. Then apply the Step SUB and transform all the available subtypes that satisfy the conditions given within the step. No particular order is necessary for applying the rest of the steps, and they can be handled in a convenient manner. However, applying the first two steps in the given order in the transformation process is essential.

## 4. THE ALGORITHM AND THE METHOD OF APPLYING IT

The text of the algorithm will be presented in *italic* to make it easy to identify them separately from the running discussion. The running discussion explains how the steps and sub steps of the algorithm are applied and a particular transformation is obtained. The final result of the transformation process will be a relational database schema (RDS) presented in the Figure 2. The RDS is a set of relations created by transforming different components: entity types, relationships, etc, of the ER model. A relation is an array of attributes sequentially and horizontally arranged. A partially created RDS will be presented at some points of the process that brings the relations created up to then.

**Prerequisite:-**

*It is required that the ER model undertaken for transforming must first be conformed to the requirements specified in the new notation scheme proposed* (Pieris, 2013).

The ER model contains three regular entity types: Employee, Department, and the Project. Regular entity types are transformed first using the Step REG as follows. Firstly the entity type Employee is transformed.

**Step REG: Transforming regular entity types**

*Follow the steps below to transform a regular entity type*

1. *Create a relation by the same name of the entity type*

   Employee[ ]

2. *Include as the primary key(PK) and the first attribute of the relation the key attribute of the regular entity type and underline it. If the entity type contains multiple key attributes, select the key attribute whose name has the first three letters similar to the first three letters of the name of the entity type.*

   Employee[<u>EmpNo</u> ]

   The entity type Employee does not have multiple key attributes but a single one, "EmpNo".

3. *Include the remaining key attributes, if existed, following the PK included and underline them separately. All the attributes must be separated by comas.*

   This sub step is skipped since no further key attributes existed in the entity type Employee.

4. *Chose the remaining non-key simple attributes of the entity type, if existed, as the remaining attributes of the relation*

Employee[<u>EmpNo</u>, Name, Address, Salary]

There are two other regular entity types: "Department" and "Project", and they should also be transformed following the same Step REG.

Department[<u>DepNo</u>, <u>Name</u>, Field ]
Project[<u>ProNo</u>, Name, Description]

Being a non-simple attribute, the attribute "Location" attached to the "Department" has not been transformed and included in the "Department" relation[2]. The two key attributes of the Department entity have been adjacently arranged in the relation complying to the requirements in the transformation step. The remaining simple attribute: "Field" should follow the key attributes.

The partially created relational database schema (RDS) is as follows

Employee[<u>EmpNo</u>, Name, Address, Salary]
Department[<u>DepNo</u>, <u>Name</u>, Field]
Project[<u>ProNo</u>, Name, Description]

Next step is to transform subtypes. In this approach, for an entity type to be a subtype it must at least have one intrinsic attribute[3] or it must participate in a relationship type with some other entity type: a regular entity type or a subtype. There are several conditions for deciding whether or not a subtype must be transformed. Sometimes certain subtypes are not necessarily to be transformed though they are acceptable subtypes.

Consider the subtype "Engineer" in the ER model. Transform it using the following step if it satisfies the conditions stated

---

[2] Department relation means the relation created by transforming the regular entity type "Department" in the ER model.
[3] Intrinsic attribute of an entity type means an attribute own by the particular entity type

### Step SUB: Transforming subtypes

*A subtype is transformed if at least one condition of the following is satisfied*

> (1). *the subtype contains an intrinsic attribute*
> (2). *the subtype lies at the N-side of a 1:N binary relationship type*
> (3). *the subtype lies at any side of a 1:1 binary relationship type, and in order to transform the relationship type, it is desired to include a PK as an FK in the relation corresponding to the subtype concerned*

Since the subtype "Engineer" satisfies the first condition, it will be transformed as follows

*Follow the steps below to transform a subtype that satisfies at least one of the above conditions.*

1. *Create a separate relation by the name of the subtype*

   Engineer[ ]

2. *Chose the PK of the relation that corresponds to the supertype of the subtype as the PK of the new relation and underline it.*

   Engineer[EmpNo]

   (Note that the supertype of the subtype is the entity type Employee, and the PK of the relation that corresponds to the entity type Employee is "EmpNo")

3. *Chose any intrinsic attribute(s) of the subtype, if existed, as the remaining attributes of the relation*

   Engineer[EmpNo, Grade]

The subtype "Manager" in Figure 1 does not have any intrinsic attribute. Thus, it does not satisfy the first condition. Though it is participated in a relationship type "Manages" the relationship type is not a 1:N relationship type. Therefore, the subtype does not satisfy the second condition as well. However, "Manages" is a 1:1 relationship type. But it is not intended to include a PK as an FK in the subtype's relation in order to transform the relationship type 'Manages" because there is another better option to perform it. Thus, the subtype "Manager" fails to satisfy even the 3$^{rd}$ condition as well. Therefore, the subtype will not be transformed.

With the transformation of the subtype "Engineer" the RDS has been changed as follows.

Employee[<u>EmpNo</u>, Name, Address, Salary]
Department[<u>DepNo</u>, <u>Name</u>, Field]
Project[<u>ProNo</u>, Name, Description]
Engineer[<u>EmpNo</u>, Grade]

Now transform the one to many (1:N) relationship type "Controls" existed in between the two regular entity types "Department" and "Project". The entity type "Project" lies at the N-side of the relationship type "Controls". Out of the two cardinality ratio value pairs: (1, 1) and (1, n) it can be said that (1, 1) to be the nearest pair to the entity type at the N-side of the relationship type and (1, n) to be the farthest.

### Step GNG: Transforming a one-to-many relationship type between two regular entity types

*Follow the steps bellow to transform a 1:N relationship type existed between two regular entity types*

1. *Chose the relation analogy[4] to the regular entity type lies at the N-side of the relationship type. Include as a FK in it following it's last attribute the PK of the non-chosen relation analogy to the other participating entity type of the relationship.*

    Project[<u>ProNo</u>, Name, Description, DepNo]

    "Project" is the entity type that lies at the N-side of the relationship type "Controls". The relation analogy to "Project" has been selected from the above RDS. "Department" is the other participating entity type, and "DepNo" is the PK of its corresponding relation which is also the non-chosen relation. Therefore, "DepNo" has been included in the "Project" relation as a FK to it. Note that the FK is included following the last attribute of the Project relation.

2. *To the including FK, give a bracketed suffix of 4 variables and set the name of the relationship type as the 1st variable.*

---

[4] A relation analogy to an entity type means a relation created transforming that entity type.

Project[ProNo, Name, Description, DepNo(Controls, , , )]

"Controls" is the name of the relationship type, and is included as the first variable in the bracket.

3. *Chose the remaining 3 variables from the values of the cardinality ratio pairs associated with the relationship type as follows*

    (i) *As the $2^{nd}$ variable, chose the min value of the cardinality ratio pair that lies nearest to the entity type at the N-side of the relationship type. (Note that the max value of the same cardinality ratio value pair is not taken for transformation. When interpreting the transformed relation, it should be assumed that this value to be 1 always.)*

    Project[ProNo, Name, Description, DepNo(Controls, 1, , )]

    The entity type that lies at the N-side of the relationship type "Controls" is Project, and the cardinality ratio pair nearest to Project is (1, 1), and its min value is 1. Note that the max value of this cardinality ratio value pair should not be transformed.

    (ii) *As the $3^{rd}$ and $4^{th}$ variables, choose the min & max values of the cardinality ratio pair that lies farthest to the entity type at the N-side of the relationship type,*

    Project[ProNo, Name, Description, DepNo(Controls, 1, 1, n)]

    The entity type that lies at the N-side of the relationship type "Controls" is Project, and the cardinality ratio pair that farthest to Project is (1, n). Thus the $3^{rd}$ and $4^{th}$ variables should be 1 and n respectively.

4. *Include any set of simple attributes, if existed, of the relationship type as attributes of the chosen relation following the FK included*

    Project[ProNo, Name, Description, DepNo(Controls, 1, 1, n)]

    No simple attributes attached to the relationship type, therefore, no attributes are appearing after the bracketed FK.

With this transformation, the RDS will be changed as follows.

Employee[EmpNo, Name, Address, Salary]

Department[<u>DepNo</u>, <u>Name</u>, Field]

Project[<u>ProNo</u>, Name, Description, DepNo(Controls, 1, 1, n)]

Engineer[<u>EmpNo</u>, Grade]

Use the following step to transform the 1:1 relationship type "Manages" existed in between the subtype "Manager" and the regular entity type "Department".

### Step SOG: Transforming a binary 1:1 relationship type between a subtype and a regular entity type

*Follow the below steps to transform the relationship type*

1. *Chose a relation corresponding to either of the participative entity types*

   Being 1:1 the relationship is symmetric between its participative entity types. This is the reason for the freedom to select either of the entity type out of the both. Thus, either the relation corresponding to the subtype "Manager" or the relation analogy to the regular entity type "Department" can be selected. However, a relation related to the subtype "Manager" has not yet been created and made available in the RDS so far created. Yet the relation corresponding to the Department has already been transformed and made available in the RDS. So we have no choice other than selecting it for the purpose.

   Department[<u>DepNo</u>, Name]

2. *Include the PK corresponding to the non-chosen relation, as an FK in the chosen relation following the last attribute of it*

   Department[<u>DepNo</u>, Name, Field, EmpNo ]

   The non-chosen relation is the relation analogy to the "Manager" subtype which is not available. According to the conditions mentioned within the Step SUB, it is not eligible to be created and will never be created. However, non-availability or non-creation of a subtype-relation is not an issue for assuming and thereby choosing its PK, since the PK of any subtype-relation is the PK of its corresponding supertype-relation. The supertype of the Manager subtype is the Employee entity type, therefore, the relation analogy to this supertype is the Employee-relation and its PK is EmpNo. Therefore EmpNo should be the PK of the "Manager" subtype relation, which is not existed and is the non chosen relation. Thus, EmpNo is included in the chosen relation "Department" following its last attribut.

Now re-consider the modified Department-relation and its last attribute EmpNo as appeared above. Since we are the creators of it, we exactly know that EmpNo to be the PK corresponding to the assumed (though not really existed in the RDS) Manager-relation. But how someone else who is going to interpret this relation should know this? Since EmpNo is the PK of the Employee relation, he might assume that Employee-relation must be the non-chosen relation (It is assumed that the respective ER model or its corresponding narrative or both are not available to anyone else). No information available within the relation. It has to be informed somehow that this particular attribute EmpNo is the PK analogy to a subtype-relation named "Manager". The method we use here is to give the name "Manager" as a prefix to this FK, EmpNo to be appeared as Manager-EmpNo. The next sub step will do this as follows.

3. *Give the FK a prefix, the name of the subtype.*

   Department[DepNo, Name, Manager–EmpNo ]

   Now the reader is forced to consider for this case that the EmpNo to be the PK of a subtype relation called Manager, and he also know that it should be the PK of Employee-relation.

4. *To the including FK, give a bracketed suffix of 4 variables and set the name of the relationship type as the 1$^{st}$ variable.*

   Department[DepNo, Name, Field, Manager–EmpNo(Manages, , , ) ]

5. *Chose the remaining 3 variables from the values of the cardinality ratio pairs associated with the relationship type as follows*

   (i). As the 2$^{nd}$ variable, chose the min value of the cardinality ratio pair nearest to the entity type analogy to the chosen relation,

   Department[DepNo, Name, Field, Manager–EmpNo(Manages, 1, , ) ]

   (ii). As the 3$^{rd}$ and 4$^{th}$ variables, choose the min & max values of the cardinality ratio pair which lies farthest to the entity type analogy to the chosen relation,

      Department[<u>DepNo</u>, <u>Name</u>, Field, Manager–EmpNo(Manages, 1, 1 ,1) ]

6. *Include any set of simple attributes, if existed, of the relationship type R as attributes of the chosen relation following the FK included*

      Department[<u>DepNo</u>, <u>Name</u>, Field, Manager–EmpNo(Manages, 1, 1 ,1), StartDate ]

After the transformation of the relationship type "Manages", the RDS is as follows.

      Employee[<u>EmpNo</u>, Name, Address, Salary]
      Project[<u>ProNo</u>, Name, Description, DepNo(Controls, 1, 1, n)]
      Engineer[<u>EmpNo</u>, Grade]
      Department[<u>DepNo</u>, <u>Name</u>, Field, Manager–EmpNo(Manages, 1, 1 ,1), StartDate ]

However, in the ER model there is another 1:1 relationship type "Assigned" existed between a subtype "Engineer" and a regular entity type "Project". Let's try to transform it using the Step SOG demonstrated above.

To apply the Step SOG.1

Being 1:1 the relationship type "Assigned" is symmetric between the two participative entity types "Engineer" and "Project". In such a situation and as specified in the Step SOG.1 any of the Engineer-relation or the Project-relation can be chosen to include the PK of the non-choosing relation. Let's see whether the Engineer-relation can be chosen. The relation is available in the RDS so far created. However, the relation corresponding to the "Project" is also available in the RDS, so that it can also be chosen. However, the Project-relation has already been complicated because of being given a suffixed FK as to transform the relationship type "Controls". On the other hand we couldn't still chose a relation analogy to a subtype to include a FK as to transform a relationship type. Thus, we would be happy to select the Engineer-relation for the purpose. Following is the transformation in brief.

Step SOG.1 gives Engineer[<u>EmpNo</u>, Grade]

Step SOG.2 gives Engineer[<u>EmpNo</u>, Grade, ProNo]

Step SOG.3 is skipped since the including FK is not a PK of a subtype-relation we skip this step

Step SOG.4 gives Engineer[<u>EmpNo</u>, Grade, ProNo(Consult, , , )]

Step SOG 5 gives Engineer[<u>EmpNo</u>, Grade, ProNo(Consult, 1,1,1)]

Finally the Step SOG 6 gives Engineer[<u>EmpNo</u>, Grade, ProNo(Consult, 1,1,1), Hours]

The new RDS is as follows.

    Employee[<u>EmpNo</u>, Name, Address, Salary]
    Project[<u>ProNo</u>, Name, Description, DepNo(Controls, 1, 1, n)]

    Engineer[<u>EmpNo</u>, Grade, ProNo(Consult, 1,1,1), Hours]
    Department[<u>DepNo</u>, <u>Name</u>, Field, Manager–EmpNo(Manages, 1, 1 ,1), StartDate ]

One might sometimes ask what would be the transformation if the Project-relation is chosen at the above Step SOG.1 instead of choosing the Engineer-relation. We will give below the transformation assuming that the Project-relation is the relation chosen instead of the Engineer-relation and we hope that the reader will be able to derive it .

Project[<u>ProNo</u>, Name, Description, DepNo(Controls, 1, 1, n), Engineer–EmpNo(Consult, 1, 1, 1), Hours]

We say that there is no issue of using the above particular transformation. However, we like the other one, so we continue with it.

The new version of the RDS is as follows

    Employee[<u>EmpNo</u>, Name, Address, Salary]
    Project[<u>ProNo</u>, Name, Description, DepNo(Controls, 1, 1, n)]
    Engineer[<u>EmpNo</u>, Grade, ProNo(Consult, 1,1,1), Hours]
    Department[<u>DepNo</u>, <u>Name</u>, Field, Manager–EmpNo(Manages, 1, 1 ,1), StartDate ]

In the ER model there is a multi-valued attribute: Location belonging to the entity type Department. Use the following Step to transform it

### Step MVA: Transforming multi-valued attributes

*Follow the following steps to transform a multi-valued attribute*

1. *Create a new relation, and name it by the name of the multi-valued attribute.*

   Location[ ]

2. *Include the PK attribute of the relation corresponding to the owner entity type of the multi-valued attribute, as the PK of the new relation and underline it*

   Location[DepNo]

3. *Include the multi-valued attribute as the second attribute and underline (The two underlined attributes will together act as the PK of the new relation while the first attribute act as a FK referring to its own relation)*

   Location[DepNo, Location]

With this transformation the RDS is changed as follows

   Employee[EmpNo, Name, Address, Salary]
   Project[ProNo, Name, Description, DepNo(Controls, 1, 1, n)]
   Engineer[EmpNo, Grade, ProNo(Consult, 1,1,1), Hours]
   Department[DepNo, Name, Field, Manager–EmpNo(Manages, 1, 1 ,1), StartDate ]
   Location[DepNo, Location]

"Dependent" is a weak entity type in the ER model with "Employee" being it's owner entity type.

The attribute: "Name" that lies in the weak entity type is a partial key attribute of it.

### Step WAK: Transforming weak entity types

*Follow the steps below to transform the weak entity type*

1. *Create a new relation by the name of the weak entity type*

2. *Include as a foreign key and as the first attribute of the relation, the PK of the relation corresponding to the owner entity type. Include the partial key attribute in the weak entity type as the second attribute of the new relation, and underline both attributes separately.*

Dependent[EmpNo, Name]

3. *Include the rest of the attributes of the weak entity type as the remaining attributes of the relation*

    Dependent[EmpNo, Name,  Sex, Relation ]

With this transformation the RDS is changed, and the final RDS is given in the Figure 2 below

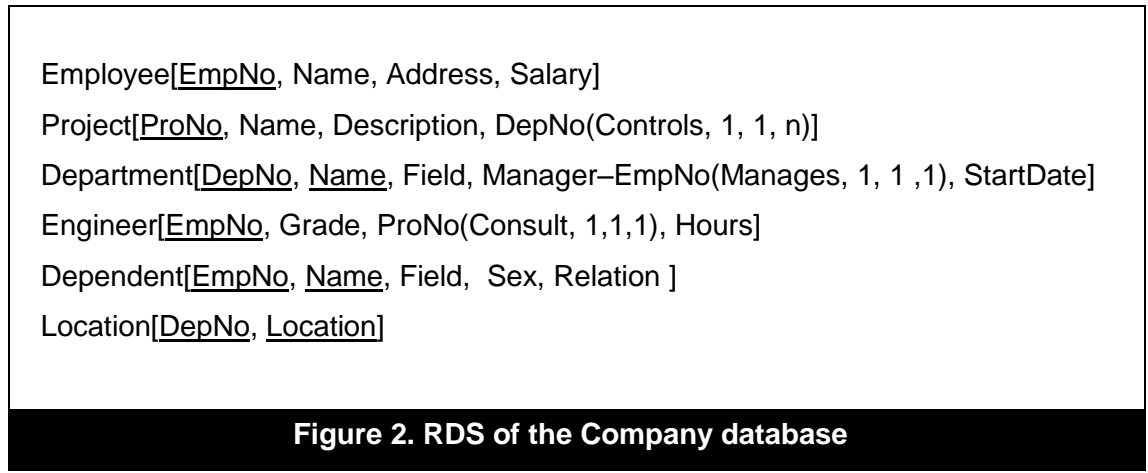

Employee[<u>EmpNo</u>, Name, Address, Salary]
Project[<u>ProNo</u>, Name, Description, DepNo(Controls, 1, 1, n)]
Department[<u>DepNo</u>, <u>Name</u>, Field, Manager–EmpNo(Manages, 1, 1 ,1), StartDate]
Engineer[<u>EmpNo</u>, Grade, ProNo(Consult, 1,1,1), Hours]
Dependent[<u>EmpNo</u>, <u>Name</u>, Field,  Sex, Relation ]
Location[<u>DepNo</u>, <u>Location</u>]

**Figure 2. RDS of the Company database**

Accordingly, 6 steps of the new algorithm have been demonstrated using an ER model given, and the ER model has also been transformed creating a RDS as given in the Figure 2 above.

## 5. FEATURES OF A RDS, ITS INTERPRETATION AND REVERTING BACK TO THE RESPECTIVE ER MODEL

We claim that this RDS has a number of qualities such as complete, accurate, non-redundant with respect to its predecessor ER model.  Since it is complete, accurate and non-redundant we further claim that it should be able to interpret and reversed back to its predecessor ER model by revoking the transformation rules applied for creating it. The transformation rules used for creating the RDS can be used again to interpret it and to reclaim the information presented in it. To facilitate this work we would like to present some common features that could be seen on a RDS and can also be derived from the same transformation rules. Most of the features presented below are followed by an example from the above RDS as to help interpret it. It has to be presumed that the RDS and the transformation rules are only available and the ER model is not known and has to be re-produced.

   i. Five types of relations can be occurred in a RDS such as 1) regular entity type relations, 2) subtype relations, 3) relationship relations, 4) multi-valued attribute relations and 5) weak entity type relations. The above RDS contains 4 types of them except the 3$^{rd}$ one: the relationship relation.
  ii. Any relation has a PK, a single attribute underlined or a combination of attributes underlined and separated by comas. Some examples for PKs are "EmpNo" in Employee and "DepNo, Location" in Location. However, a combination of underlined attributes at the beginning does not necessarily to be a PK of a relation, for example, "DepNo, Name" in Department.
 iii. The PK of a regular entity type relation is always of a single attribute. It must be underlined and be placed at the beginning of the relation. Its name must be similar to the name of the relation by first three letters, examples: EmpNo in Employee, ProNo in Project, and DepNo in Department. Therefore, Employee, Project and Department are regular entity type relations in the RDS. Each of them must have a corresponding regular entity type in the respective ER model.
  iv. The PK of a relation analogy to a regular entity type can also be included as the PK of a new relation with no suffixes or prefixes to any of the attributes in the relation. If the new relation contains
       a. Only one underlined attribute then it should be analogy to a subtype whose supertype is the regular entity type. For example Engineer is a subtype relation and also the Engineer is a subtype of the supertype Employee,
       b. Two attributes which are the only underlined ones in the relation then the relation must be analogy to a multi-valued attribute whose owner to be the regular entity type. The second attribute of the relation should be the multi-valued attribute, and the name of the relation should be the name of the multi-valued attribute. For example, Location is a multi-valued attribute relation,

- c. More than two attributes with two of them are underlined then the relation must be analogy to a weak entity type whose owner to be the regular entity type, for example, Dependent is a weak entity type relation.
v. Foreign keys (FK) in a relation have bracketed suffixes. No FK can be existed without a bracketed suffix (Some FKs may have prefixes). Examples for FKs are the DepNo in Project, Manager-EmpNo in Department, etc.
vi. A FK is sometimes given with a prefix, and the prefix indicates the owner-a subtype-of the PK that the FK refers. The ownership of the PK is shared in between the regular entity type and the subtype. Such a subtype may or may not have a separate relation in the RDS. An example for this kind of FK is "Manager-EmpNo" which is in Department. Accordingly, there may or may not be a subtype relation "Manager" in the RDS, but there must be a subtype "Manager" in the ER model. The supertype must be the Employee entity type.

Note that further interpretation of FKs and their given bracketed suffixes and there by implementing the revealing information in the ER model should be done by revoking the transformation rules used for creating those FKs and their suffixes. We say that by this exercise the RDS can be fully interpreted and the exact ER model can be re-established.

## 6. CONCLUSION

The existing ER to relational transformation algorithm is not able to provide a straightforward transformation of an ER model, and it also fails to transform most of the semantic information of a ER model created. We have presented a new transformation algorithm in order to resolve the problem. However, the ER model must first be drawn following a certain standard (cite). We argue that the current approach can transform an ER model to the relational model completely, accurately and semantically clearly, and the resulted schema can also be reversed straightforwardly to the original ER model without being distorted. To test the validity of this claim the RDS must first be interpreted. This interpretation can be done by re-applying the algorithm on the RDS but with a view to revoke the steps already performed as to create it. As to support this interpretation process we have derived some common feature of the produced RDS(they must be common to any RDS produced in this approach). They can be used as a help interpret the RDS and re-create the respective ER model again. This paper presents only a part of the algorithm, and we wish to present the rest soon. We further think that our notion of revoking the steps of the algorithm performed in creating the RDS must also be formalized. Accordingly, we have already created a reverse transformation algorithm, and in the future we will present it as well.